\documentclass[twocolumn, 10pt]{wmrDoc}

\usepackage{fourier}
\usepackage[T1]{fontenc}
\usepackage{hyperref}
\hypersetup{colorlinks,linkcolor={black},citecolor={black},urlcolor={black}}
\usepackage{graphicx}
\usepackage[font=small,labelfont=bf]{caption}
\hypersetup{colorlinks=true,urlcolor=black}
\usepackage{tabularx}
\newcolumntype{Y}{>{\centering\arraybackslash}X}
\renewcommand{\figurename}{Figure}
\usepackage{titlesec}
\titleformat{\section}
  {\normalfont\fontsize{12}{15}\bfseries}{\thesection}{1em}{}

\title{NewsTweet: A Dataset of Social Media Embedding in Online Journalism}

\author{
    \textbf{Munif Ishad Mujib$^\dagger$, 
    Hunter Scott Heidenreich$^\dagger$, 
    Colin J. Murphy$^\dagger$}\\
    \textbf{Giovanni C. Santia$^\dagger$, 
    Asta Zelenkauskaite$^\ddagger$, 
    and Jake Ryland Williams$^\dagger$}
    \affiliation{
        Drexel University\\
        College of Computing and Informatics$^\dagger$, 
        Communication Culture and Media$^\ddagger$\\
        {\{mim52, hsh28, cjm486, gs495, az358, and jw3477\}@drexel.edu}
    }
}

\begin{document}
\maketitle

\begin{abstract}
    \it
    The inclusion of social media posts---tweets, in particular---in digital news stories, both as commentary and increasingly as news sources, has become commonplace in recent years. In order to study this phenomenon with sufficient depth, robust large-scale data collection from both news publishers and social media platforms is necessary. This work describes the construction of such a data pipeline. In the data collected from Google News, 13\% of all stories were found to include embedded tweets, with sports and entertainment news containing the largest volumes of them. Public figures and celebrities are found to dominate these stories; however, relatively unknown users have also been found to achieve newsworthiness. The collected data set, NewsTweet, and the associated pipeline for acquisition stand to engender a wave of new inquiries into social content embedding from multiple research communities.
\end{abstract}

\section{Introduction}
The appearance of user-generated content from social media in web-based news articles is rapidly becoming a familiar phenomenon to readers. Studies have been conducted on the impact of including such content on reader perception. As a novel and engaging method of surfacing the voice of the general population as well as a more authentic channel for reporting statements from sources of news, be they persons or organizations, this phenomenon has engendered interest in multiple research communities.

Due to this interest in social media content \emph{embedding}, as referred to in this work, in terms of sourcing routines in the news~\cite{paulussen2014} and understanding social media sourcing in particular~\cite{lecheler2016}, a need for pertinent data collection and sharing is apparent. 

The presented data set, ``NewsTweet'', is an ongoing automatic collection from a broad range of news content categories. It collects user-level activity in addition to selected embedded tweets that may be used for predictive modeling. Further, it includes the proportions of tweets and associated users, thus providing a multilateral approach to social media sourcing.

The construction of this data set is the first necessary step in the study of the social media embedding phenomenon: the design of a custom data stream spanning news outlets and genres that sit at the critical intersection of edited news and noisy social content. 

\section{Methodology}
Social media content from a variety of platforms appears in web articles in embedded form. In terms of reported embedding volume share, the largest platforms are Twitter (59\%), YouTube (26\%), Instagram (14\%), and Facebook (1\%)~\cite{sam:SocialEmbeds}. An exhaustive data collection pipeline should capture, follow, and analyze the embedded content from all of these sources. However, differing platform standards limit deep access to their content. As such, NewsTweet, in its first version presented here, collects social data only from Twitter, which has become the de-facto standard for social media analysis. Twitter also remains the largest source of embedded social content. The data collection described in this work exhibits expected proportions of embedded content from different platforms (see Table~\ref{tab:embeddings}).

\begin{table}[h!]
    \centering
    \small
    \renewcommand{\arraystretch}{1.4}
    \begin{tabular}{lrr}
    \hline
    \multicolumn{1}{c}{{\bf Platform}} & \multicolumn{1}{c}{{\centering \bf Articles}} & \multicolumn{1}{c}{{\bf Embeddings}}\\\hline
    Twitter & 39,498 (57.33) & 92,299 (68.05)\\
    YouTube & 19,557 (28.39) & 27,960 (20.61)\\
    Instagram & 8,021 (11.64) & 13,241 (9.76)\\
    Facebook & 1,785 (2.59) & 2,081 (1.53)\\
    Reddit & 27 (0.039) & 32 (0.023)\\
    TikTok & 2 (0.0029) & 17 (0.0013)\\\hline
    \bf{Total} & 68,890 (100.0) & 135,630 (100.0)\\\hline 
   \end{tabular}
   \caption{
    Volumes of embedded content by platform being embedded. 
    The column {\bf Articles} exhibits the numbers (and percentages) of articles 
    that contain embeddings from the platforms, 
    while {\bf Embeddings} indicates the total number of embeddings (and percentages by them) observed from each.
    }
   \label{tab:embeddings}
\end{table}

\subsection{RSS Tracking}
NewsTweet attempts to acquire data on the interaction of embedded content with news production while maintaining representation of the various existing content categories in news. With this stated goal, a reasonable starting point for data collection was deemed to be a generalized news aggregator, specifically, the Google News RSS feeds~\cite{GoogleRSS}. These feeds offer a common breakdown of news into 8 sections: Business (B), Entertainment (E), Health (H), Nation (N), Sports (S), Technology (T), World (W), and Headlines (X)\footnote{An aggregation within the aggregator.}. Each of theses is accessed by requesting along the following URL archetype from the \texttt{https://news.google.com} domain:
\begin{center}
    \small{\texttt{/news/rss/headlines/section/topic/$\langle$SECTION$\rangle$}}
\end{center}
The result of each request is a collection of hyperlinks to news articles along with limited other metadata. All newly-seen hyperlinks are then accessed for their HTML content, which includes embedded social media content. The resulting HTML content is processed to find and extract embedded tweets when present.

\subsubsection{Caveats}
RSS feed data collection has several caveats. The Google News RSS feeds lack official documentation. However, help forums discussing usage can be found. Google offered a software application, Google Reader, from 2005 to 2013 that allowed users to follow and read RSS feeds. Google Reader was shut down due to declining user numbers, signaling a shift in policy and product direction within the company regarding RSS feeds. While the feeds continue to be available---presumably, as part of the Google News product---Google may choose to sunset their availability at any time. However, since RSS is an open standard and other RSS news aggregators are and will foreseeably continue to be available, collection can be readily switched to a different aggregator. Moreover, the surfeit of news sources already uncovered by this collection effort could be directly monitored, creating new custom RSS feeds. However, Google News commands a substantial volume of users due to its close integration with Google Search~\cite{trielli2019search}. As such, the selection of articles aggregated on this platform is uniquely significant. Thus, while the loss of the Google News RSS feeds would certainly have a substantial impact to analyzing online news-related phenomena, alternatives for data collection will remain available.

A particular artifact of the Google News feeds is the inclusion of YouTube video pages as articles. During the data collection process, it was discovered that $\sim$6.8\% of articles were YouTube pages. Although the HTML content of these pages is downloaded, they are ignored in subsequent analyses, since no social media embeddings can appear on a YouTube page. 

\begin{figure}[t]
    \centering
    \includegraphics[width=0.99\columnwidth]{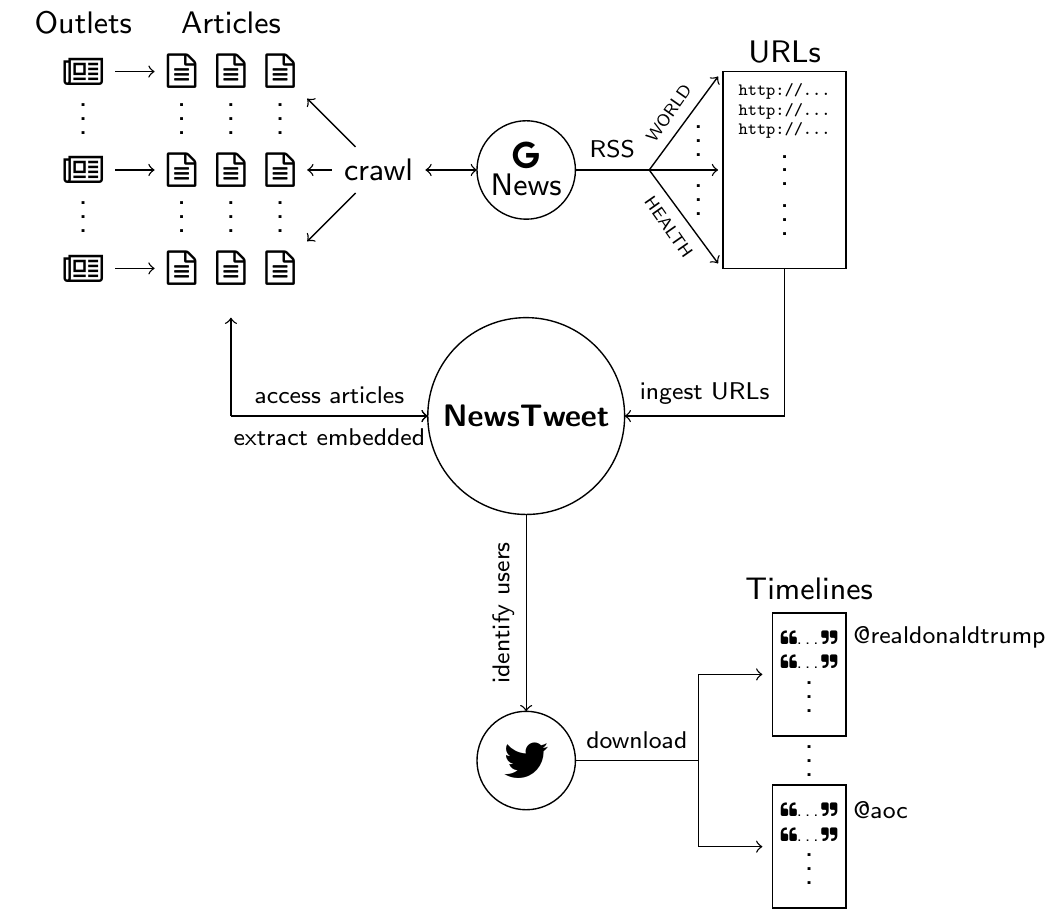}
    \caption{
     Schematic diagram of the NewsTweet data collection pipeline.
    }
    \label{fig:acquisition}
\end{figure}

\subsection{Twitter Data Acquisition}
Embedded tweets, which primarily display user handles, timestamps, tweet text, and attached media to readers, appear in articles with limited metadata. From this metadata, individual tweet IDs are used to construct calls to the Twitter API for full tweet objects. These objects include user profile data and often geographic information as well. Once full tweet objects have been downloaded, the user IDs are leveraged to direct mass data collection from the Twitter platform and access large, continuous portions of the users' timelines (i.e., their tweets in chronological order). These timelines are accessed in reverse chronological order and restricted to the users' most recent 3,200 tweets according to Twitter API policies.

\begin{table}
    \centering
    \resizebox{1.0\columnwidth}{!}{
    \renewcommand{\arraystretch}{1.4}
    \begin{tabular}{lrrrrr}
        \hline
        \multicolumn{1}{c}{\centering \bf \S} & \multicolumn{1}{c}{\bf Article} & \multicolumn{1}{c}{\bf Embedded} & \multicolumn{1}{c}{\bf Embeds} & \multicolumn{1}{c}{\bf Tweets} & \multicolumn{1}{c}{\bf Users}\\
        \hline
		{\bf B} & 41,006 & 2,285 (6\%) & 4,488 & 3,510 (78\%) & 2,314\\
		{\bf E} & 49,263 & 6,827 (14\%) & 17,233 & 13,380 (78\%) & 9,152\\
		{\bf H} & 12,498 & 228 (2\%) & 378 & 351 (93\%) & 283\\
		{\bf N} & 38,461 & 4,353 (11\%) & 8,783 & 7,032 (80\%) & 3,778\\
		{\bf S} & 59,553 & 14,429 (24\%) & 35,841 & 27,857 (78\%) & 9,398\\
		{\bf T} & 34,435 & 3,105 (9\%) & 5,118 & 3,782 (74\%) & 2,313\\
		{\bf W} & 24,531 & 2,119 (9\%) & 3,704 & 3,057 (83\%) & 1,864\\
        {\bf X} & 14,152 & 1,872 (13\%) & 3,969 & 3,540 (89\%) & 2,267\\\hline
        {\bf A} & 273,899 & 35,218 (13\%) & 79,514 & 60,523 (76\%) & 27,838\\
        \hline
    \end{tabular}
    }
    \caption{
    Descriptive statistics by section (\S) for four months of data collection. \S\textbf{A} indicates all sections combined.
    Columns indicate total {\bf Articles}, number of articles {\bf Embedded},
    the total number of {\bf Embeds} present, 
    the number of unique {\bf Tweets}, and {\bf Users} embedded.
    Percentages indicate the share of articles with embeds for that section and the proportion of unique tweets for that section. 
    }
    \label{tab:embedrates}
\end{table}

\begin{table*}[t]
    \centering
    \renewcommand{\arraystretch}{1.4}
    \resizebox{1.00\textwidth}{!}{
      \begin{tabular}{ccccccccc}
        \hline
        {\bf WORLD} & {\bf BUSINESS} & {\bf HEADLINES} & {\bf HEALTH} & {\bf TECHNOLOGY} & {\bf SPORTS} & {\bf NATION} & {\bf ENTERTAINMENT} & {\bf ALL} \\
        \hline
        realDonaldTrump, 464 & realDonaldTrump, 263 & realDonaldTrump, 399 & jameelajamil, 10 & FortniteGame, 131 & wojespn, 871 & realDonaldTrump, 810 & WWE, 171 & realDonaldTrump, 2150 \\
        rcmpmb, 35 & elonmusk, 206 & elonmusk, 59 & realDonaldTrump, 7 & UniverseIce, 123 & ShamsCharania, 343 & AOC, 148 & realDonaldTrump, 134 & wojespn, 895 \\
        JZarif, 25 & PopeyesChicken, 36 & AOC, 36 & jencurran, 7 & PokemonGoApp, 57 & FOXSoccer, 304 & atrupar, 72 & LilNasX, 97 & ShamsCharania, 352 \\
        atrupar, 25 & charlieburns, 31 & atrupar, 31 & CDCgov, 6 & NintendoAmerica, 55 & JohnOwning, 183 & justinamash, 68 & KimKardashian, 93 & FOXSoccer, 326 \\
        IreneSans, 23 & Josh\_Rager, 26 & wojespn, 23 & EcoWatch, 5 & wongmjane, 38 & TheSteinLine, 176 & ABC, 43 & misscp, 72 & elonmusk, 301 \\
        \hline
        ChiefManak, 0.11 & ToaAz, 0.1 & SuperASASSN, 0.17 & KevinH\_PhD, 0.33 & Cr8Beyond, 0.07 & KSTiLLS, 0.08 & BlindDensetsu, 0.08 & bethisloco, 0.08 & BlindDensetsu, 0.1 \\
        shannongsims, 0.17 & PopeyesChicken, 0.11 & quantumpenguin, 0.2 & jacionline, 0.33 & Raf\_\_\_m, 0.09 & NevadaFootball, 0.11 & NWSSanDiego, 0.14 & JonxDanyy, 0.08 & PopeyesChicken, 0.11 \\
        Laurie\_Garrett, 0.17 & alieward, 0.11 & councilofdc, 0.25 &  ssteingraber1, 0.5 & fortrisen, 0.11 & krisnoceda, 0.12 & NYPDCT, 0.14 & neiltyson, 0.09 & alieward, 0.11 \\
        Southcom, 0.2 & BlindDensetsu, 0.12 & Peter\_Grindrod, 0.25 & BorisJohnson, 0.5 & googlephotos, 0.11 & ElisaraEdwards, 0.12 & EllenLWeintraub, 0.14 & Yo\_Bala, 0.09 & ChiefManak, 0.11 \\
        Andrew\_Denney, 0.2 & Macys, 0.17 & WilliamShatner, 0.27 & metoffice, 0.5 & paul\_irish, 0.12 & RFootball, 0.14 & rap30, 0.17 & HOT97, 0.11 & fortrisen, 0.11 \\
        \hline
        MeAndVan, 4 & charlieburns, 9 & JoaquinCastrotx, 2.5 & thrasherxy, 2 & coiiiiiiiin, 3 & livvalice, 4 & WPXICropper, 5 & keanuwtm, 14 & legitbecky, 21 \\
        DrGetahun, 3 & Morehouse, 3 & \_Carabinieri\_, 2 & metoffice, 0.5 & JeffBezos, 2 & Kenny\_Kangaroo, 4 & Morehouse, 3 & ChristiesInc, 7 & keanuwtm, 14 \\
        billmaher, 2 & LombardiHimself, 2 & BillNye, 2 & ssteingraber1, 0.33 & milanmilanovic, 2 & AZDesertSwarm, 3 & LaurenDake, 3 & Mike\_Dougherty, 5 & charlieburns, 9 \\
        dsquintana, 2 & bylaurenfox, 1.67 & AustinScottGA08, 2 & jacionline, 0.33 & mariokarttourEN, 2 & NHRA, 3 & RepMattSchaefer, 3 & MikePosner, 4 & jonvoight, 8 \\
        NWSBirmingham, 1 & diamondnagasiu, 1.25 & haya2e\_jaxa, 1.67 & Maricopahealth, 0.25 & Jokereed, 2 & TheCaveman316, 3 & VPPressSec, 2 & \_DavidGilmour, 4 & WillieMcNabb, 8 \\
        
        \hline
      \end{tabular}
    }
    \caption{
      Descriptive statistics exhibiting the top five 
      most-embedded users (by total embeds, top section),
      most re-embedded users (by fraction of embeddings unique, ascending), and
      most effectively-embedded accounts (by number of tweets embedded 
      out of their total number produced over their embedding period).
    }
    \label{tab:topusers}
\end{table*}

Once the initial timeline depth is accessed for a given user, their account is then added to a running list of users whose timelines are tracked and ``topped off'', i.e., new tweets are gradually added to the original collection. Thus, once a tweet from a user is found embedded in an article, that user is subsequently continuously tracked for new tweets. Since the number of users to follow rapidly grows beyond the limits of the Twitter API's standard free tier, users are randomly sampled at regular intervals for new tweet downloads or top-offs to ensure that all users are eventually reached. This random sampling ensures that user timeline data collection does not lag behind the real-time stream. Efficient maintenance of the stream will ultimately require an automated queueing system that prioritizes topping off more active users more often. These enhancements are essential and remain a priority for implementation, as one of the goals of this project is to release this acquisition software to the community to advance research activity in this area. A full schematic of the data collection pipeline is presented in Figure~\ref{fig:acquisition}. 

\section{Descriptive Statistics and Observed Patterns}
Data collection initiated on May 15th, 2019. As of September 11th, 2019, the stream resulted in the acquisition of 273,899 articles (2,302 articles per day, on average). 35,218 (13\%) of these had embedded tweets (296 articles per day, on average)\footnote{However, the number of articles with embeds from all platforms is nearly double this, and is quantified in Table~\ref{tab:embeddings}.}.

\subsection{Embedding Patterns and News Sections}
The articles containing embedded social content are not uniformly distributed over the eight Google News sections. This can be seen in Table~\ref{tab:embedrates}, which presents these and other descriptive statistics by section. This table showcases that overall, 13\% of articles included embeddings. Yet, content categories showed variation in proportions of embeddings. The Sports category featured the highest proportion of the embedded content  (24\%) of the articles, followed by Entertainment (14\%) with the smallest one being Health amounting to only 2\%. A reversed pattern is seen in terms of unique tweets embedded (93\% in Health vs. 78\% in Sports).

\subsection{Embedding Patterns and Users}
Many users are repeatedly embedded. Among these, @realdonaldtrump~\footnote{Users are treated as public figures and their usernames are presented. No messages are associated with the presented usernames in the hope of reducing potential risk of harm from public exposure.} is a clear outlier, having been embedded an order of magnitude more in the Nation and World sections. Top embedded users by total embeds by section are shown in Table~\ref{tab:topusers}. These are consistently newsworthy accounts. When users are sorted by the lowest fraction of unique tweets, we see a different group of users in the middle section of Table~\ref{tab:topusers}. For these users, only a few of their unique tweets were found newsworthy, but were embedded repeatedly by journalists. Thus, we can see from this view that a completely different set of users -- often celebrities, and well-known organizations -- receive highly focused attention, perhaps around very specific, time-limited events.

Taking a different view towards the effectiveness of users at getting embedded, we again see a completely different picture. In the bottom section of Tab.~\ref{tab:topusers}, the number of unique tweets each user had embedded is divided by the number of tweets they produced in their period of being embedded. Here, perhaps an unexpected picture emerges, as the users who received the most embeds for the least tweets, actually received more unique embeds than they produced unique tweets. This is possible only because there is a lag between the times when a tweet is produced and embedded. For these users, journalists had to 'catch readers up' on the stories that emerged from possibly unknown individuals. Thus, we consider if some users had a back story that necessarily had to be told in order for their newsworthiness to be qualified. Under this hypothesis, we ask if these effectiveness top-ranking users represent those  who truly gained celebrity status from their embeddings, and perhaps come from a less known status before.

\begin{table}[h!]
    \renewcommand{\arraystretch}{1.4}
    \centering
    \resizebox{1.0\columnwidth}{!}{
    \begin{tabular}{ccc}
      \hline
          {\bf Rank} & {\bf Total Articles} & {\bf Average Embeds}\\
          \hline
	  1 & foxnews.com, 10,038 & blavity.com, 9.73 \\
	  2 & cnn.com, 8,930 & milehighreport.com, 6.39 \\
	  3 & reuters.com, 4,640 & baltimoreravens.com, 5.87 \\
	  4 & nytimes.com, 4,497 & cornnation.com, 5.53 \\
	  5 & espn.com, 4,324 & teenvogue.com, 5.08 \\
	  6 & cnbc.com, 4,154 & dodgersdigest.com, 4.60 \\
	  7 & usatoday.com, 3,935 & chepicap.com, 4.57 \\
	  8 & nypost.com, 3,866 & titansonline.com, 4.38 \\
	  9 & washingtonexaminer.com, 3,732 & nba.nbcsports.com, 4.28 \\
	  10 & thehill.com, 3,503 & junkee.com, 4.23 \\
	  \hline
    \end{tabular}
    }
    \caption{
      Top-ranked domains by article count and average number of embedded tweets per article.
    }
    \label{tab:topsites}
\end{table}

\subsection{Embedding Patterns and Mass Media}
The news side of the data collection also presents interesting characteristics. At the coarsest level, Google News RSS feeds covered 5,961 different domains in all. Table~\ref{tab:topsites} presents the top-ranked domains sorted both by total articles and average embedded tweets per article.\footnote{For the average embeddings per article ranking, an ad-hoc threshold is applied to exclude domains that have published fewer than 10 articles. This prevents domains that have few articles but embed an abnormal number of tweets from dominating the ranking.}

The top hosts that appear in Table~\ref{tab:topsites} are what might be expected for both metrics. Top hosts by total articles are well known mass media organizations, whereas top hosts by average embeddings seem to largely belong to categories that have higher embedding rates, i.e., Sports and Entertainment (further observable in the results in Table~\ref{tab:embedrates}).

\section{Applications and Future Work}
This data set stands to inform and enable numerous angles of inquiry into the phenomenon of social media content embedding. The patterns immediately visible in the collected data pose intriguing questions. In particular, this data connects well with existing research into user commenting typology based on the contribution practices in news to gauge patterns of influence~\cite{zelenkauskaite2017}. 

A particularly salient question is that of newsworthiness. What makes social media content newsworthy? Which users are deemed more newsworthy? What is the likelihood for a user's content to be selected for embedding? What role does the user's status on the social platform (verification, affiliation, and reach) play on their newsworthiness? When social media content is embedded in the news, how often is the content itself the source of news, and how often is it commentary on news? What defines ``staying power'' in the news? How are ``influencers'' and internet celebrities born? What effects do sudden newsworthiness bring to a user's social media following and behavior? 

Many scholars of the early days of social media, such as~\cite{hermida2010}, emphasized the potential of social media to ``break through from a one-way, asymmetric model of communication to a more participatory and collective system, where citizens have the ability to participate in the news production process.'' How much of this potential has been achieved? To what degree and when are `regular' people embedded into news stories? Answers to many of these questions can potentially be achieved through exploration of and experimentation using NewsTweet data.

By collecting timeline data from historically newsworthy users, an opportunity to study content production and selection also arises. Since the nature of news volatile and stories evolve over time, NewsTweet data can be utilized to potentially construct meta-stories. These can both be long-term phenomena such as the emergence of the COVID-19 pandemic and associated social frenzy over a number of months as well as shorter-lasting, focused events such as the compromise and abuse of high-profile accounts in a coordinated cryptocurrency scam in July 2020.

Some authors~\cite{zelenkauskaite2018} have suggested that users have very limited pre-defined spaces and journalistic gatekeeping remains the key element in the user-generated content process. The collected data and illustrated data collection pipeline present an opportunity to implement novel theoretical and methodological approaches to analyze social media integration in the news. While social media in journalism has solicited academic interest, most approaches have so far focused on content analysis of the articles themselves, and have only had access to relatively-small scale and manually collected data sets. The NewsTweet data set's description and its framework for a data collection pipeline is shared with the broader interdisciplinary research community in the hope of advancing new avenues of scholarship in social media, digital media, and journalism. 

\section*{Acknowledgments}
This document is based upon work supported by the National Science Foundation under grant no. \#1850014.

\bibliographystyle{wmrDoc}
\bibliography{newstweet}

\begin{thebibliography}{1}

\bibitem{paulussen2014}
Paulussen, S., and Harder, R.~A., 2014.
\newblock ``Social media references in newspapers''.
\newblock {\em Journalism Practice, {\bf 8}}(5), pp.~542--551.

\bibitem{lecheler2016}
Lecheler, S., and Kruikemeier, S., 2016.
\newblock ``Re-evaluating journalistic routines in a digital age: A review of
  research on the use of online sources.''.
\newblock {\em New media \& society, {\bf 18}}(1), pp.~156--171.

\bibitem{sam:SocialEmbeds}
SAM, 2019.
\newblock The state of social embeds.
\newblock
  \url{https://cdn.samdesk.io/static-content/The-State-of-Social-Embeds.pdf}.
\newblock Accessed: Aug. 20th, 2019.

\bibitem{GoogleRSS}
Google, 2018.
\newblock How to get an {RSS} feed from {G}oogle {N}ews in 2018.
\newblock
  \url{https://support.google.com/googlenews/forum/AAAAKvAM41IWV2z_hgA6p0/?hl=en&gpf=%23!topic%2Fnews%2FWV2z_hgA6p0}.
\newblock Accessed: Aug. 20th, 2019.

\bibitem{trielli2019search}
Trielli, D., and Diakopoulos, N., 2019.
\newblock ``Search as news curator: The role of {G}oogle in shaping attention
  to news information''.
\newblock In Proceedings of the 2019 CHI Conference on Human Factors in
  Computing Systems, ACM, p.~453.

\bibitem{zelenkauskaite2017}
Zelenkauskaite, A., and Balduccini, M., 2017.
\newblock ````{I}nformation {W}arfare'' and online news commenting: Analyzing
  forces of social influence through location-based commenting user
  typology.''.
\newblock {\em Social Media+ Society, {\bf 3}}(3).

\bibitem{hermida2010}
Hermida, A., 2010.
\newblock ``From tv to twitter: How ambient news became ambient journalism''.
\newblock {\em Media/Culture Journal, {\bf 13}}(2).

\bibitem{zelenkauskaite2018}
Zelenkauskaite, A., 2018.
\newblock ``Value of user-generated content: Perceptions and practices
  regarding social and mobile media in two italian radio stations''.
\newblock {\em Journal of Radio \& Audio Media, {\bf 25}}(1), pp.~23--41.

\end{thebibliography}
\end{document}